# A single-source molten salt synthesis of uniform octahedral $Na_2Ti_3O_7$ particels composed of nanorods

Huan Xie,[#a] Jin Chen,[#a] Bingyu Lei[a] and Lei Zhou*[ab]

[a] Department of Biomedical Engineering, School of Life Science and Technology, Huazhong University of Science and Technology, Wuhan, 430074, P. R. China.

[b] Advanced Biomaterials and Tissue Engineering Center, Huazhong University of Science and Technology, Wuhan, 430074, P. R. China.

[#] These authors contributed equally to this work.

[*] Corresponding Author

E-mail: lei.zhou.public@gmail.com

Tel: +86-27-87792216

**Abstract**

$Na_2Ti_3O_7$, as one of the most potential anode electrode materials, is expected to play an important role in various fields. In this work, we synthesized uniform hollow octahedral $Na_2Ti_3O_7$ particles through directly sintering a kind of precursors obtained by $(NH_4)_2TiF_6$ and $NH_4OH$ in the presence of sodium-containing cationic polyacrylamide. XRD, SEM, TEM were used to characterize precursors and $Na_2Ti_3O_7$ particles. In addition, the synthesis conditions and the formation mechanism were discussed in detail. This single-source molten salt method is flexible, efficient and controllable. Moreover, the strategy of this method makes it possible for us to synthesize other sodium titanate materials.

## 1. Introduction

In recent years, sodium-ion batteries have attracted extraordinary attention owing to their low cost and raw materials in abundance. They offer a promising scalable energy storage alternative to current lithium-ion batteries (LIBs) [1-3]. As a kind of anode electrode materials, sodium titanates, having the general formula of $Na_2Ti_nO_{2n+1}$ ($2 \leq n \leq 9$), exhibit high ionic exchange ability and high $Na^+$ ion conductivity [4,5]. $Na_2Ti_3O_7$ (sodium trititanate), a kind of $Na_2Ti_nO_{2n+1}$ (n=3), has been proved to be a very potential anode material with low voltage plateau (around 0.3 V) [6,7], high charge capacity [8,9] and stable storage capacity [3,10]. It has received considerable attention in various fields such as sodium-ion batteries [4,11,12], ion exchangers [10,13], photocatalysts [14], bioactive ceramics [15] and sensors [16]. Up to now, some methods have been reported for synthesis of $Na_2Ti_3O_7$ crystals, such as solid-state molten salt reaction [1,6,7,13,14], sol-gel reaction [3,17], hydrothermal reaction [9,12,18]. However, there are some shortages in these methods, such as the long sintering time for solid-state molten salt reaction (10 h ~20 h), the long sol drying time for sol-gel reaction, the high temperature (and a Teflon-lined autoclave is essential) for hydrothermal reaction. These shortages limit the massive production of $Na_2Ti_3O_7$. In order to synthesize $Na_2Ti_3O_7$ with high crystallinity more efficiently, an efficient and flexible method is demanded.

In our previous work, we synthesized rod-like $Na_2Ti_6O_{13}$ particles through a single-source molten method [19]. This method is flexible and efficient, which just needs to directly sinter a kind of precursors obtained by $(NH_4)_2TiF_6$ and $H_3BO_3$ in the presence of sodium salts. Due to the similarities of crystal structure between $Na_2Ti_6O_{13}$ and $Na_2Ti_3O_7$ [20], this single-source molten salt approach may be applied to synthesize $Na_2Ti_3O_7$ particles, only if we can successfully prepare suitable precursors.

In this study, we synthesized uniform hollow octahedral $Na_2Ti_3O_7$ particles through directly sintering a kind of precursors obtained by $(NH_4)_2TiF_6$ and $NH_4OH$ in the presence of sodium-containing cationic polyacrylamide. The hollow octahedral $Na_2Ti_3O_7$ particles are composed of nanorods with high crystallinity. In addition, the formation process and the reaction conditions of this single-source molten salt method were discussed. In comparison with the conventional methods, this method is more flexible, efficient and controllable. Moreover, the strategy of this method also makes it possible for us to synthesize other sodium titanate materials.

## 2. Experiment

### 2.1 Materials and methods

**Synthesis of precursors:** Sodium-containing cationic polyacrylamide (labeled as Na-CPAM) (Tianjin Zhiyuan Chemical Reagent) and $(NH_4)_2TiF_6$ (Sigma-Aldrich) were dissolved in 50 ml distilled water, then $NH_4OH$ solution (25 wt%) was added drop by drop with stirring. The final concentrations of $(NH_4)_2TiF_6$, $NH_4OH$, Na-CPAM were 0.5 M, 1.0 M and 4 wt%, respectively. After the addition of $NH_4OH$ solution, the mixture solution became turbid and was treated at 25 ºC for 30 min. The precipitates were collected and separated by centrifugation, washed (×3) with distilled water and dried at 35 ºC for 12 h.

**Synthesis of $Na_2Ti_3O_7$:** The furnace was preheated to 800 ºC. Then the precursor powders were directly put into the furnace and sintered at 800 ºC for 2 h.

### 2.2. Characterization

Scanning electron micrographs (SEM) and EDX spectrum were obtained using a FEI Nova Nano SEM 450. Powder X-ray diffraction (XRD) studies were performed using a PANalytical X'Pert PRO diffractometer. Transmission electron micrographs (TEM) were obtained using a FEI Tecnai G2 F30.

## 3. Results and discussion

A kind of stable precursors with octahedral morphology was synthesized in the hydrolysis system of $(NH_4)_2TiF_6$ and $NH_4OH$ with adding sodium-containing cationic polyacrylamide (labeled as Na-CPAM). SEM images and XRD patterns of the samples prepared by adding different concentrations of Na-CPAM are shown in Figure 1 and Figure 2, respectively. When the Na-CPAM was not added in the hydrolysis system (0 wt% Na-CPAM), the products are rod-like particles (Figure 1a), and the XRD result (Figure 2a) confirms the particles are pure $(NH_4)_2TiOF_4$. With adding Na-CPAM, a kind of octahedral particles was synthesized. Under low

Na-CPAM concentration (< 4 wt%), the octahedral particles coexist with rod-like $(NH_4)_2TiOF_4$ particles (Figure 1b and 1c). With increasing the Na-CPAM concentration, the number of octahedral particles increases, together with the decreasing of rod-like $(NH_4)_2TiOF_4$ particles. Especially, when the Na-CPAM concentration is higher than 4 wt%, the products are almost octahedral particles (Figure 1d and 1e), and the peaks of $(NH_4)_2TiOF_4$ (JCPDS-49-0161) phase are almost wiped out (Figure 2d and 2e). These SEM and XRD results demonstrate that a kind of octahedral precursors was synthesized.

To identify the component of octahedral precursors, EDX spectrum was carried out. As shown in Figure 1f, the octahedral precursor particle contains N, F, Na, Ti, O elements. According to our previous work, $Na^+$ can partly substitute the position of $NH_4^+$ in $NH_4TiOF_3$ [19]. Similarly, in this system, $Na^+$ coming from Na-CPAM could partly substitute the position of $NH_4^+$ in $(NH_4)_2TiOF_4$ and form the precursors. In addition, Na-CPAM, as a cationic surfactant, is crucial to stabilize the structure of precursor particles, which can bond with the inorganic species and control the growth rate of various faces [21]. Since, it seems very probable that the chemical formula of precursor is $(NH_4)_xNa_yTi_zO_\beta F_{x+y+4z-2\beta}$.

To obtain pure $Na_2Ti_3O_7$, the furnace was preheated to 800 ºC before sintering. Then the precursor powders were directly put into furnace for sintering at 800 ºC for 2 h. After sintered, the white powders were characterized by XRD and SEM. The XRD pattern (Figure 3a) indicates that the products are pure $Na_2Ti_3O_7$. SEM and HRSEM image of a $Na_2Ti_3O_7$ particle are displayed in Figure 3b and 3c. After the sintering process, the previous overall octahedral shape remains, although the surface is rougher (Figure 3b). HRSEM image of a $Na_2Ti_3O_7$ particle confirms that the hollow octahedral particle is constructed by large nanorods (Figure 3c) with a length of 3~5 μm and a diameter of ~300 nm. The details of the $Na_2Ti_3O_7$ nanorods were further investigated using TEM and SAED. The results are given in Figure 3d. It confirms the $Na_2Ti_3O_7$ nanorods are highly crystalline, which are consistent with conventional molten salt methods [1,6,7,13,14]. These results indicate that the single-source molten salt method is a highly efficient strategy to synthesize hollow octahedral $Na_2Ti_3O_7$ particles composed of nanorods. Furthermore, the hollow octahedral structure may enhance the contact area between the electrode and electrolyte and can improve the efficiency of electrolyte penetration [2,13]. Therefore, the hollow octahedral $Na_2Ti_3O_7$ particles obtained in this method may have a potential application in battery field.

For the single-source molten salt method, sintering time is very important. In order to investigate the effect of sintering time, the as-prepared precursor powders were sintered at 800 ºC for different times, and corresponding XRD patterns are displayed in Figure 4. At the early stage of sintering (1 h), peaks of $TiO_2$ (rutile phase and anatase phase) appear. This evidence suggests that the $Ti^{4+}$ from precursor particles convert to $TiO_2$ and the $Na^+$ form precursor particles may convert to some amorphous sodium salts, which is consistent with our previous work [19]. With the sintering process proceeding, $TiO_2$ and amorphous sodium salts convert to $Na_2Ti_3O_7$. The $Na_2Ti_3O_7$ particles obtained in this method are thermodynamic stable without any generation of impurity for longer sintering time (> 3 h). In this process, it is worth noting that the amorphous sodium salts formed at the early stage are essential for the subsequent molten salt reaction. To prove it, the powders prepared by sintering precursor at 800 ºC for 1 h were washed with deionized water for three times (amorphous sodium salts were almost eliminated), and later sintered at 800 ºC for 2 h. The XRD pattern shows that the products are rutile phase (Figure 4e) without any evidence of $Na_2Ti_3O_7$. SEM images of samples during the sintering process are shown in Figure 5. At the beginning of the sintering process, the previous smooth surface of octahedral shape becomes rough due to the formation of $TiO_2$ particles and amorphous sodium salts (Figure 5a). When the precursors were sintered at 800 ºC for 2 h, the stable hollow octahedral shape $Na_2Ti_3O_7$ particles composed of nanorods were formed (Figure 5b). However, long sintering time (over 2 h) would lead to the collapse of octahedral shape (Figure 5c and 5d), which could be caused by excessive crystal growth of nanorods. This result suggests that the sintering time is crucial for final component and particle morphology of products.

Moreover, the sintering temperature, as the driving force for phase transformation and crystal growth, is another vital factor for the single-source molten salt reaction. In order to further investigate the effect of sintering temperature, the as-prepared precursor powders were sintered at different temperatures (400 °C ~700 °C) for 2 h. The corresponding XRD patterns and SEM images are displayed in Figure 6 and Figure 7, respectively. In comparison with 800 °C samples (Figure 3), it is obvious that lower temperature cannot drive the formation of $Na_2Ti_3O_7$. For the XRD patterns of samples obtained at 400 °C ~600 °C, only peaks of anatase phase particles appear (Figure 6a, 6b and 6c). Even at 700 °C, the XRD pattern shows mixed peaks of anatase and rutile (Figure 6d). SEM images show that the octahedral precursor particles just become rougher (Figure 7a) or collapse into some nanoparticles (Figure 7b-7d) at low sintering temperatures (< 800 °C) condition. These results indicate that 800 °C is essential for molten salt reaction between $TiO_2$ and amorphous sodium salts.

Besides the sintering process used above, we also tried another sintering process. Contrary to preheat the furnace to 800 °C before sintering, the precursor was put into furnace first. Then the precursor powders were heated from room temperature to the required 800 °C and sintering at 800 °C for 2 h. The XRD pattern (see Figure 8a) indicates that the products obtained by this sintering process are composites of $Na_2Ti_3O_7$ and $Na_2Ti_6O_{13}$. SEM (Figure 8b) and HRSEM (Figure 8c) image show that the previous overall octahedral shape of precursor particles cannot maintain and collapse into some random rods. The result indicates that the first sintering process is better to prepare pure octahedral $Na_2Ti_3O_7$ particles composed of nanorods.

With all results and discussions described above, the process of single-source molten salt reaction can be summarized in Figure 9. At the early sintering stage, the octahedral shape precursor particles, as the single reactant, would convert to $TiO_2$ and amorphous sodium salts. With the sintering process proceeding, a molten salt reaction occurs between $TiO_2$ and amorphous sodium salts to form $Na_2Ti_3O_7$. During the whole process, the previous octahedral shape maintains and the final $Na_2Ti_3O_7$ particles are composed of nanorods.

## 4. Conclusions

In this work, we synthesized a kind of stable octahedral precursors in the hydrolysis system of $(NH_4)_2TiF_6$ and $NH_4OH$ with adding Na-CPAM. After a sintering process, the octahedral precursor particles convert to hollow octahedral $Na_2Ti_3O_7$ composed of nanorods through a single-source molten salt reaction. The sintering condition and the formation mechanism of $Na_2Ti_3O_7$ were discussed. In comparison with conventional approaches, this approach is more flexible, efficient and controllable. Moreover, the $Na_2Ti_3O_7$ particles obtained in this method may have a potential application in battery field due to the hollow octahedral overall structure, and the strategy of this method makes it possible for us to synthesize other sodium titanate materials.

## 5. Acknowledgment

This work is supported by the Fundamental Research Funds for the Central Universities (HUST, Grant No. 2014QN120 and 2016YXMS259) and the National Natural Science Foundation of China (NSFC, Grant No. 51172084). The authors also thank the Analytical and Testing Center of Huazhong University of Science and Technology (HUST) for providing SEM, TEM, AFM and XRD measurements.

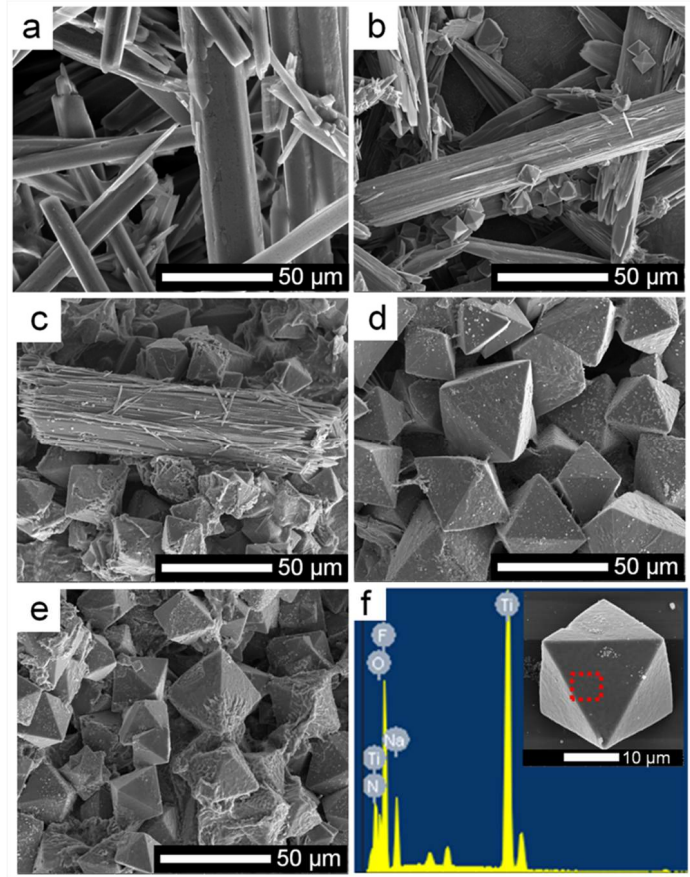

**Figure 1** SEM images of samples obtained by adding different concentrations of Na-CPAM. (a) 0 wt%; (b) 1 wt%; (c) 2 wt%; (d) 4 wt%; (e) 8 wt%. (f) EDX result and HRSEM image of samples obtained by adding 4 wt% Na-CPAM.

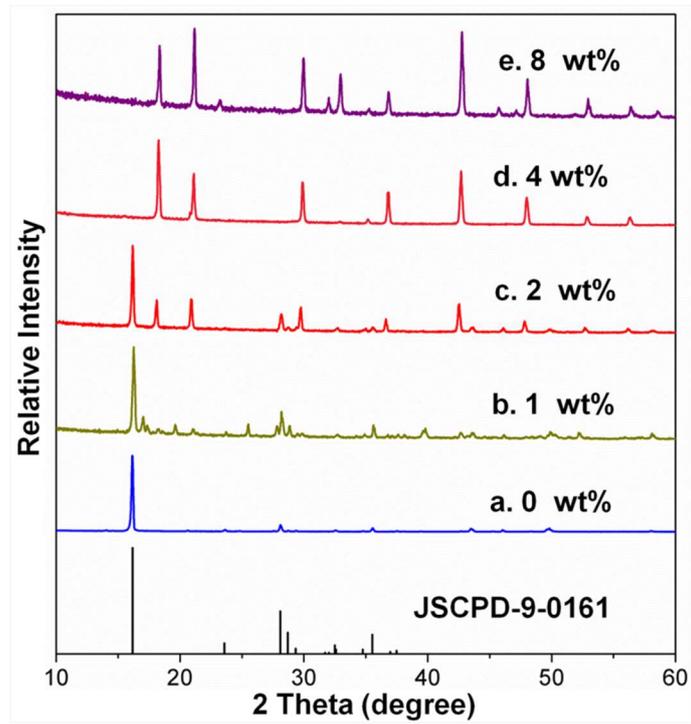

**Figure 2** XRD patterns of samples obtained by adding different concentrations of Na-CPAM. (a) 0 wt%; (b) 1 wt%; (c) 2 wt%; (d) 4 wt%; (e) 8 wt%.

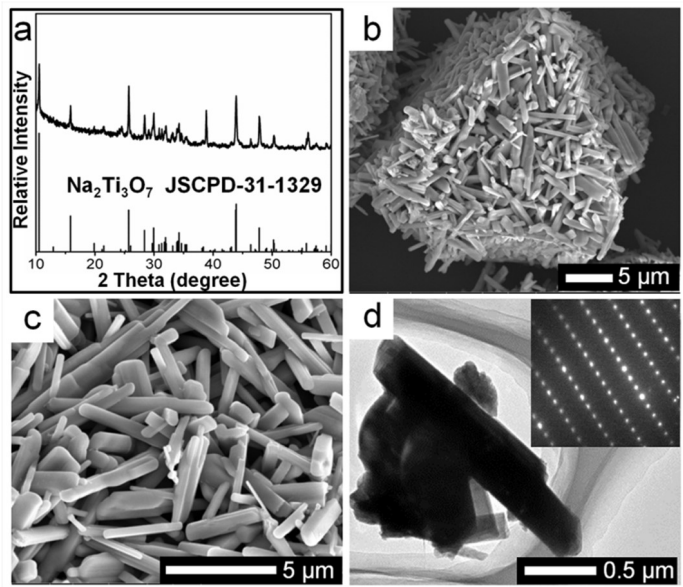

**Figure 3** XRD pattern (a), SEM image (b), HRSEM image (c) and TEM image (d) of samples obtained by sintering precursor powders at 800 ºC for 2 h. The inset is the SEAD pattern.

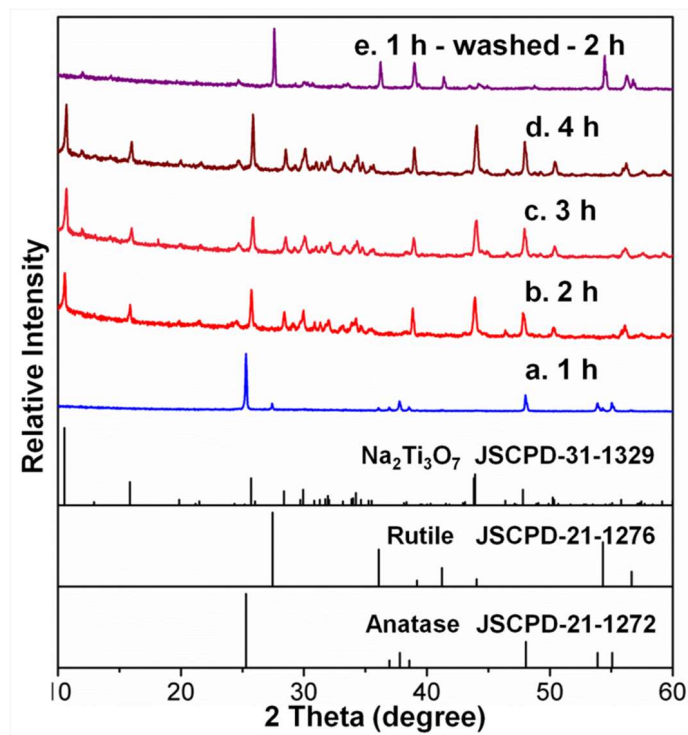

**Figure 4** XRD patterns of samples obtained by sintering precursor powders at 800 °C for different times. (a) 1 h; (b) 2 h; (c) 3 h; (d) 4 h; (e) After sintering precursor powders at 800 °C for 1 h, the powders were washed by deionized water (3 times), and later sintered at 800 °C for 2 h.

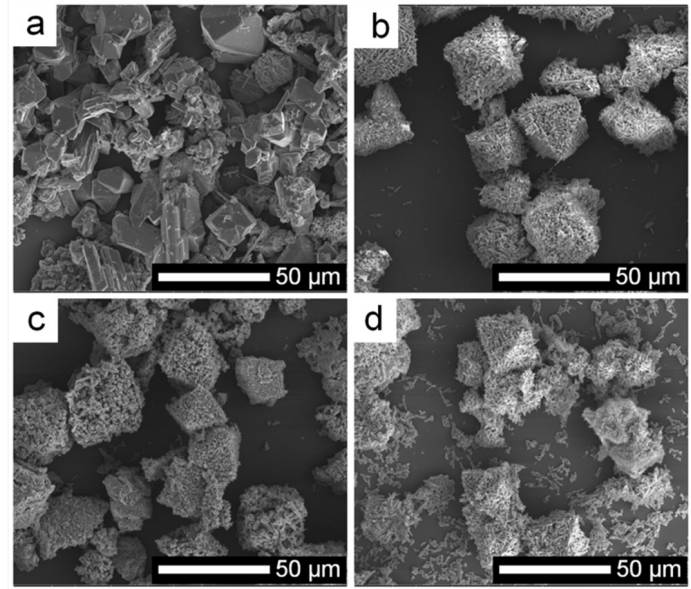

**Figure 5** SEM images of samples obtained by sintering precursor powders at 800 °C for different times. (a) 1 h; (b) 2 h; (c) 3 h; (d) 4 h.

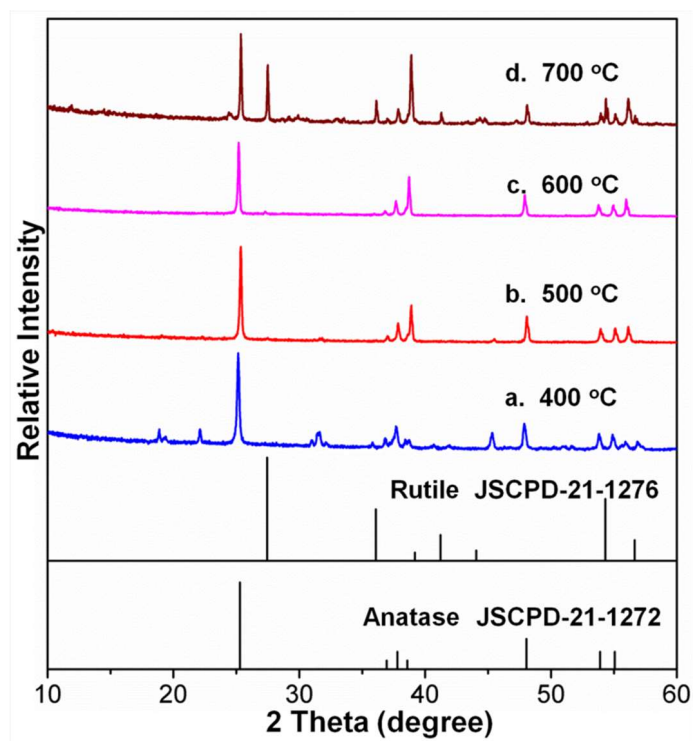

**Figure 6** XRD patterns of samples obtained by sintering precursor powders at different temperatures for 2 h, (a) 400 ºC; (b) 500 ºC; (c) 600 ºC; (d) 700 ºC.

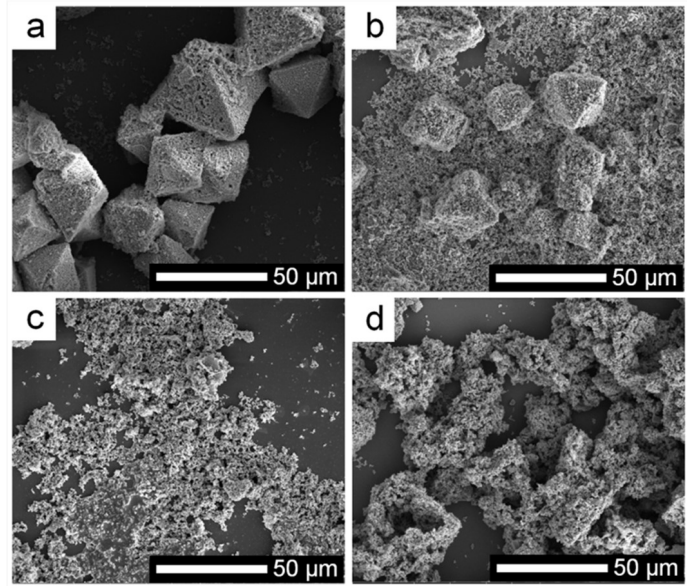

**Figure 7** SEM images of samples obtained by sintering precursor powders at different temperatures for 2 h, (a) 400 ºC; (b) 500 ºC; (c) 600 ºC; (d) 700 ºC.

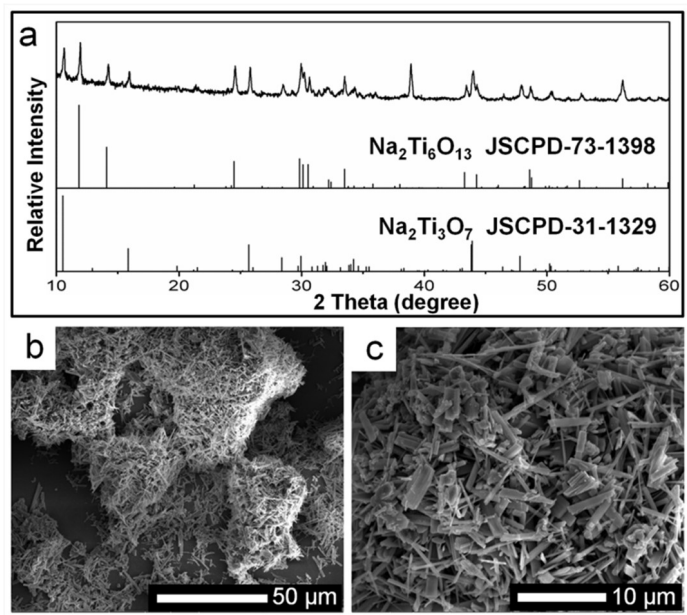

**Figure 8** XRD pattern (a), SEM image (b) and HRSEM image (c) of samples obtained by slowly sintering precursor powders at 800 °C for 2 h.

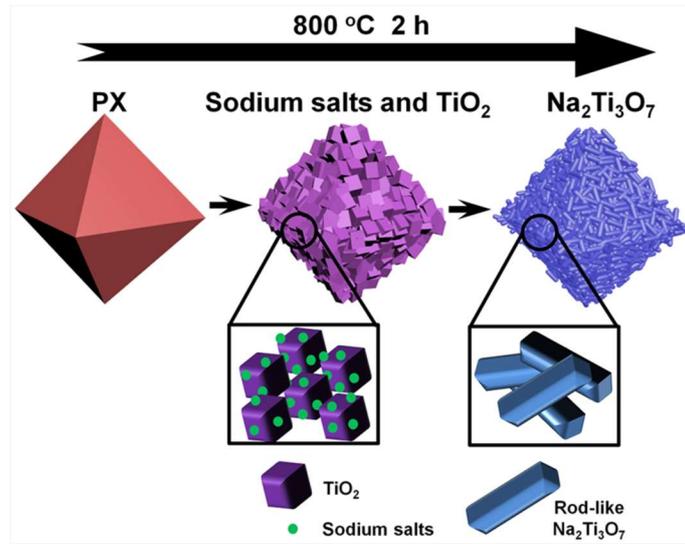

**Figure 9** The formation mechanism of the Na$_2$Ti$_3$O$_7$ synthesized through the single-source molten salt method.